\pdfoutput=1
\documentclass{article}
\usepackage[utf8]{inputenc}
\usepackage[T1]{fontenc}
\usepackage{lmodern}
\usepackage{amsmath}
\usepackage{amssymb}
\usepackage{graphicx}
\usepackage{booktabs}
\usepackage{array}
\usepackage{float}
\usepackage{url}
\usepackage{cite}
\usepackage[margin=1in]{geometry}
\usepackage[hidelinks]{hyperref}
\hypersetup{pdftitle={Ghostline: a six-node Wi-Fi CSI motion tracker with node-to-node Links. System description and a pilot evaluation},
            pdfauthor={Sebastian Cochinescu, Ionut Moldovanu, Alexandru Constantin}}
\graphicspath{{figures/}}
\title{Ghostline: a six-node Wi-Fi CSI motion tracker with node-to-node Links.\\ System description and a pilot evaluation}

\author{
Sebastian Cochinescu\\
University of Bucharest, Bucharest, Romania\\
\texttt{sebastian.cochinescu@drd.unibuc.ro}
\and
Ionuț Moldovanu\\
Independent researcher, Bucharest, Romania\\
\texttt{robopitorg@gmail.com}
\and
Alexandru Constantin\\
Military Technical Academy ``Ferdinand I'', Bucharest, Romania\\
\texttt{alex@devmotion.ro}
}

\date{13 October 2026}

\begin{document}

\maketitle

\begin{abstract}
Ghostline is a device-free motion tracking system built on six ESP32-S3 Wi-Fi boards and one
consumer router. The boards' Wi-Fi driver reports channel state information (CSI) for every
received frame with the sender's address, and a modified open-source firmware forwards both, so
every board carries a CSI stream for the router and for each other board: seven radios give up to
36 directed Links over 21 Spans instead of a star of six. Each Link is compared with a
self-forming quiet Baseline, disturbed Links vote on a floor grid, and the grid becomes anonymous
Tracks. The system was built and run during a three-day hackathon in an ordinary room. We describe
it in full and report a pilot evaluation with one walker: spoken cues at seven marked positions, a
two-minute free walk and five minutes of empty room, synchronised to video within 0.18 s (median).
Walking between marks, the Track lay a median 0.85 m from the straight path and 1.56 m (95\% CI
1.29 to 2.83) from a cue-based reference trajectory (constant speed between cues, not measured
positions), with 46\% of walk updates within 2 m of it; the free walk was tracked in 94\% of
updates. The limits: a standing person is lost within seconds, a false Track appeared in 15.9\% and
2.9\% of updates in two empty-room periods, and informal runs that afternoon did not separate two
walkers. Two offline replays through the same pipeline show where this comes from: with the
node-to-node Links removed and settings unchanged, the tracked share of walk updates falls from
90\% to 19\% and the free walk from 96\% to 60\%; a plain radio tomographic image on the same per-Span
evidence, at the same false rate, matches the grid and tracker at 2 m (48\% against 47\%) and is
ahead at 1 m (36\% against 21\%). On this recording the Tracks depend on the Link set, and the
imaging stage adds no measurable accuracy. We state what the pilot establishes and what it does
not.
\end{abstract}

\section{Introduction}

A body that crosses a radio link shadows and scatters the signal, and a receiver that reports
its channel can see it. Device-free passive localisation built on this effect predates Wi-Fi
sensing as a field \cite{youssef2007dfp}. Radio tomographic imaging (RTI) turned it into a method: a mesh of static radios,
one received-signal-strength link per pair, and a linear inverse problem that places the
attenuation on a floor grid \cite{wilson2010rti}. Channel state
information (CSI) from commodity Wi-Fi chipsets then gave each link per-subcarrier amplitude and
phase instead of one strength number, and a decade of work used it for presence, activity,
gesture and localisation \cite{yang2013rssi2csi,ma2019csisurvey}. Low-cost ESP32
boards expose CSI from firmware and bring the sensor cost down to that of a microcontroller
board \cite{hernandez2020esp32,hernandez2023edge,atif2020wiesp}.

Most commodity-board deployments keep the topology of the Wi-Fi network they ride on: an access
point transmits, each board receives, and N boards give N links that all start at one point. RTI
draws its resolution from links that cross the area from many directions. Ghostline brings the
two together on cheap hardware. The ESP32 Wi-Fi driver delivers CSI for every received frame
together with the sender's MAC address \cite{espidf_csi}, so a small firmware change lets each board also report the frames it hears from
the other boards: six boards and a router give up to 36 directed Links over 21 Spans in one
room. The system compares each
Link with a self-forming quiet Baseline, lets the disturbed Links vote on a floor grid in a
metric map of the room, and turns the grid into anonymous Tracks: a moving point with a number,
no camera, no wearable, no identity.

This paper is a pilot evaluation, and we say so. The system was built during the three-day
European Defense Tech Hackathon in Bucharest (9 to 11 October 2026), on earlier single-board
tests with RuView, and evaluated on the hackathon's last afternoon with one walker in one room. We describe the system from firmware to Tracks, the
evaluation protocol, the measured numbers and the failures, and we separate measured evidence
from design hypotheses throughout. The hypothesis that motivates the link set, that node-to-node
Links track better than the router-only star, is tested on the pilot recording by an offline
replay with the node-to-node Links removed (Section~\ref{sec:e1}); it is one session in one room, so the
result is evidence for the layout tested, not a general effect size.

Contributions:

\begin{enumerate}
\def\labelenumi{\arabic{enumi}.}
\item
  A complete description of a six-node deployment in which seven radios sustain 36 directed
  CSI Links, from firmware to Tracks, built on an open-source server and firmware (Sections~\ref{sec:system}
  and~\ref{sec:census}).
\item
  A cued-walk protocol with spoken cues, logged cue times and a phone video synchronised to the
  cue log within a median 0.18 s, so that every tracker output can be checked against where the
  walker was (Section~\ref{sec:eval}).
\item
  Measured one-walker results in a 48~m$^2$ room: a median 0.85 m from the walked path between
  marks and 1.56 m from a cue-based reference trajectory, 94\% tracked during a free walk, and
  the failure modes of standing still, two walkers and empty-room false Tracks (Sections~\ref{sec:results}
  and~\ref{sec:discussion}).
\item
  Two controlled offline comparisons on the same recording: with the node-to-node Links
  removed and the settings unchanged, the tracked share of walk updates falls from 90\% to 19\%
  and the free walk from 96\% to 60\% (Section~\ref{sec:e1}); a plain RTI reconstruction on the same
  per-Span evidence is indistinguishable from the grid and tracker at the 2 m scale and ahead
  at 1 m (Section~\ref{sec:rti}).
\item
  An explicit boundary between what the pilot shows and what it does not (Section~\ref{sec:limitations}), and the
  offline experiments the recordings permit (Section~\ref{sec:planned}).
\end{enumerate}

\section{Related work}\label{sec:related}

Device-free localisation through radio links is twenty years old, and the mesh formulation that
Ghostline follows is RTI. RTI models each link's extra attenuation as the integral of a loss
field along an ellipse around the line of sight and solves for the field on a grid with
regularised least squares \cite{wilson2010rti}. Later RTI work moved from mean attenuation to
its variance, which tracks motion through walls \cite{wilson2011vrti}, to a kernel distance
between RSS histograms that sees stationary and moving people \cite{zhao2013kernel}, and to
multiple-target tracking on the image \cite{bocca2014mtt}. RTI has also been fed with CSI from
MIMO measurements instead of RSS \cite{xu2018csirti}. Ghostline keeps the geometric idea, a floor grid that every
Span votes on, but takes its per-Link evidence from CSI rather than from received signal
strength, and forms its Baseline continuously rather than from a one-off empty-room capture. Section~\ref{sec:pipeline} gives the statistic, the
Baseline estimator and the voting rule.

CSI-based Wi-Fi sensing mostly runs on laptop or router chipsets with a handful of links, and
its localisation branch spans fingerprinting, model-based and learned approaches
\cite{yang2013rssi2csi,ma2019csisurvey,zafari2019survey}; its device-free tracking systems
typically work from one or two links with Doppler and angle estimates, from IndoTrack
\cite{li2017indotrack} to the recent CoTrack, which tracks across rooms from a single bistatic
link with a median trajectory error of 1.14 m \cite{wang2026cotrack}. The edge branch of the
field, surveyed by Hernandez and Bulut, asks what can run on a microcontroller-class board and
what signal processing survives the move \cite{hernandez2023edge}.
Ghostline sits in that branch: the boards only report CSI, all processing runs on one laptop,
and the tracker uses signal processing alone, with no trained model, so that it needs no
training data from the room.

The closest systems to ours are those in which several radios serve as both transmitters and
receivers, so that the link set grows with the square of the number of radios. That is the RTI
sensor network, with dedicated low-power radios and RSS \cite{wilson2010rti,bocca2014mtt}, and
it now exists on ESP32 boards with CSI: Khamaisi and Rodrigues ran a nine-node ESP32-C3 mesh
with 72 sensing links for twelve days in a home and found, for motion detection, that a single
well-placed link outperformed the fused mesh, a dilution effect from links whose Fresnel zones
miss the activity \cite{khamaisi2026lessismore}; OpenCSI exposes each link of an eight-node
ESP32 mesh as a Z-score against its own quiet-period mean and standard deviation, learned from a
short empty-room bootstrap and updated online in later quiet periods, and transfers a binary occupancy threshold across rooms and chip
generations \cite{khamaisi2026opencsi}. An all-pairs CSI mesh on commodity boards is therefore
not new. What Ghostline adds is not the mesh but what is done with it: the Links are used for
localisation and tracking on a floor grid rather than for detection or occupancy, the probe
traffic is the boards' own uplink and the router's traffic with a per-transmitter rate cap and
no added sensing transmissions, the per-Link quiet level is a different estimator, the 10th percentile of
the motion statistic over the last minute rather than a quiet-period mean and standard deviation
(how Ghostline's Baseline behaves in a rearranged room was not tested), and the
evaluation ties every tracker output to a synchronised video and tests the Link set and the
imaging stage by replay. The dilution finding
of Khamaisi and Rodrigues concerns detection by a classifier; whether it carries over to
localisation, where crossing links are the source of resolution, is exactly what the replay of
Section~\ref{sec:e1} asks, for one room. ESP32 CSI toolkits \cite{hernandez2020esp32,atif2020wiesp}
collect CSI on several boards with the access point as the only transmitter. The open-source
RuView project \cite{ruview} supplies the firmware base and a server with its own presence and
people estimate, which Ghostline shows beside its Tracks but never merges into them.

\section{The Ghostline system}\label{sec:system}

\subsection{Hardware and Links}

Six ESP32-S3 boards (16 MB flash) and one TP-Link Archer AX53 router form the sensor network, on
2.4 GHz Wi-Fi channel 1. Each board runs RuView 0.8.12 firmware with our modifications and sends
CSI frames over UDP to a laptop on the router's LAN, at most 8 per second per transmitter. The ESP-IDF CSI
callback delivers, with every received frame, the CSI and the sender's MAC address
\cite{espidf_csi}; the firmware forwards both, so a board reports a separate
CSI stream for the router and for each of the other five boards. A directed Link is one transmitter-to-receiver path with its own CSI stream;
the two Links between the same pair of radios form one Span. Six boards and a router give up to
36 directed Links over 21 Spans.

The firmware is RuView 0.8.12 with two additions. Each board holds a peer table with the MAC
addresses of the other boards, and it emits a second datagram format, CSI frame v2: a 40-byte
little-endian header followed by one int8 I/Q pair per subcarrier. The header carries the
receiving node id, the transmitter's node id (1 to 6 for a board, 254 for the router, 0 for an
unknown transmitter) and the transmitter's MAC address, the 802.11 sequence number, the
frame-control byte, RSSI, noise floor, channel frequency, a mesh timestamp when valid, and a
per-receiver frame counter whose gaps count drops. Frames from transmitters that are neither the
router nor a listed peer are dropped on the board, before a per-transmitter rate limit of 8
frames per second, so a receiver emits at most about 48 v2 frames per second; the original v1
frame continues for the router Link only, so the unmodified RuView server still sees one Link per
board. No sensing transmissions are added: the frames measured are the boards' own UDP uplink and
the router's traffic. On the laptop the transmitter of a v2 frame is resolved from its MAC
address against the inventory, because a board's peer table can be stale after a reflash, and the
id in the header is used only when the MAC is unknown. The firmware was built reproducibly from
the upstream 0.8.12 archive in a pinned ESP-IDF container image whose digest is recorded with the
build.

Anchor positions (router and every board) are measured once at install in a metric map frame
whose origin is a floor corner of the room, z up. The room itself is a 3D scan placed in the same
frame, which also supplies the walls and obstacles the tracker and the evaluation use.

\subsection{From Links to Tracks}\label{sec:pipeline}

On the laptop a relay forwards every datagram to the unmodified RuView server and to our
tracker. The tracker keeps, for every Link, a motion statistic and a quiet Baseline that forms by
itself from the Link's own recent history; a short Hold-still, when everyone in the room keeps
still, seeds the Baseline quickly, and a Baseline can be saved by name and reloaded for the same
room, anchors and channel. A Link whose statistic rises above its Baseline is disturbed. The
disturbed Spans vote on an occupancy grid over the floor, and the grid's peaks become
detections that a tracker turns into Tracks, each with an id, a position, an uncertainty radius
and an age. The tracker publishes ten updates per second over a WebSocket to a browser page that
draws the room, the anchors, the live Links and the Tracks. Every live datagram is also written
to a black-box recording with the settings and events of the session, so that a session can be
replayed through the current pipeline offline.

The pipeline, in order, with the parameter values used in the scored session (the defaults of
the pipeline configuration, unchanged except the Track cap of 1):

\begin{enumerate}
\def\labelenumi{\arabic{enumi}.}
\item
  Link features. For every frame the per-subcarrier amplitude becomes log(1 + \textbar h\textbar).
  Always-zero subcarriers (DC and guards) and subcarriers 0 and 1 are dropped. The mean is
  subtracted per 64-subcarrier LTF block (router frames carry three blocks, peer frames two),
  which cancels gain steps and the block-to-block gain change that a single per-frame mean
  leaves in the feature and that reads as motion. Duplicate frames are dropped; a Link that
  switches subcarrier count for 50 frames re-locks and takes a fresh Baseline.
\item
  Motion statistic. The temporal standard deviation of each normalised subcarrier over the
  Link's window, averaged over subcarriers. The window holds 6 frames at the Link's measured
  rate, clamped to 0.6 to 2.0 s, and is fixed when the Baseline is taken; a quality from 0 to 1
  is how full the window is, and fewer than 3 frames give no new estimate.
\item
  Baseline. The 10th percentile of the motion statistic over the last 60 s, with samples above
  3 times the current level never entering, so a Link only has to be quiet for 10\% of the last
  minute. A Hold-still of a few seconds seeds it (each Link needs 10 frames; the scored
  session's saved Baseline came from a 20 s Hold-still), a Link's first 2 s give an automatic
  Baseline, and a saved Baseline can be loaded instead.
\item
  Room level. The ratio of each Link's statistic to its Baseline is divided by the 0.4 quantile
  of all live Links' ratios once that quantile exceeds 1.5, so that one person who disturbs most
  Links in a small reflective room does not remove the grid's contrast.
\item
  Disturbance. d = 1 - exp(-(ratio - 1.4) / 1.0), clipped to 0 to 1, with instant attack and a
  release no faster than 0.5 s; it decays when frames stop, and a Link silent for 3 s leaves its
  Span.
\item
  Spans and grid. Evidence enters once per Span as the quality-weighted mean of its live Links'
  Disturbance, with weight the sum of their qualities: two votes for a healthy pair, one for a
  one-way Span, none when starved. The grid covers the room polygon's bounding box in 0.25 m
  cells masked to the polygon minus obstacles. Each Span's zone is a Gaussian around its segment
  with a 1-sigma half-width of 0.7 m at mid-span, narrower at the anchors; Spans crossing an
  obstacle are scaled by 0.5. Per cell, evidence = sum over Spans of q w (d - 0.35 (1 - d)),
  clipped at 0, so quiet Spans through a cell count against it. Span Disturbances are smoothed
  with an exponential moving average (alpha 0.5 per 10 Hz step), and the image is divided by a
  running maximum that never falls below 1 and decays with an 8 s time constant, so that a
  quiet room stays dark.
\item
  Detection. The strongest cell is taken, the Disturbance it explains is removed from every Span
  through it, and the step repeats up to 4 times; picks are at least 1.5 m apart and above
  max(0.3, 0.5 times the first pick); the position is the centroid of the nearby bright cells.
\item
  Tracks. One constant-velocity Kalman filter per Track (process noise 0.6~m/s$^2$, measurement
  noise 0.9 m divided by the square root of the detection strength), Hungarian assignment on the
  squared Mahalanobis distance, gated at 13.8 (99.9\%) and by a distance of 2.5 m that grows by
  1.5 m/s without a detection, up to 4.0 m. A
  confirmed Track without a detection takes the brightest cell in its gate as a soft measurement
  if it is at least 0.1; soft measurements never keep a Track alive. A manoeuvre detector raises
  the process noise at turns and stops (by 4 per unit of excess, up to 16 times). A Track is
  confirmed after 3 consecutive associations, needs a detection strength of 0.8 to be born and
  must be 3.5 m from a confirmed Track; it coasts 3 s without detections (5 s once older than
  4 s), two confirmed Tracks within 1.2 m merge into the older id, and no new Track is born while
  the cap is reached. The published position is the state propagated 0.5 s ahead, a lead that
  fades over 1 s without a real detection.
\end{enumerate}

The pipeline is signal processing only. No model is trained, and no data from the room is needed
beyond the anchor positions and the Baseline.

\subsection{Room check}

Before a session an empty-room check confirms that the Baseline is quiet: it reports the share of
Links that still read motion and whether any Track appears while the room is known to be empty.
The check runs over a timed window and passes when at least three live Links with a Baseline are
present in at least 80\% of the window, the mean share of Links reading motion is at most 10\%,
and a Track is present in at most 5\% of updates; it lists the Links that moved in at least 30\%
of the window. The live health display calls the room busy when, over the last 30 s, more than
half of the live Links were disturbed on average; in that state the grid has no quiet Links to
vote against and Tracks are not trustworthy.

\section{Pilot evaluation: one walker, one room}\label{sec:eval}

\subsection{Room and anchors}

\begin{figure}[H]
\centering
\includegraphics[width=0.72\textwidth]{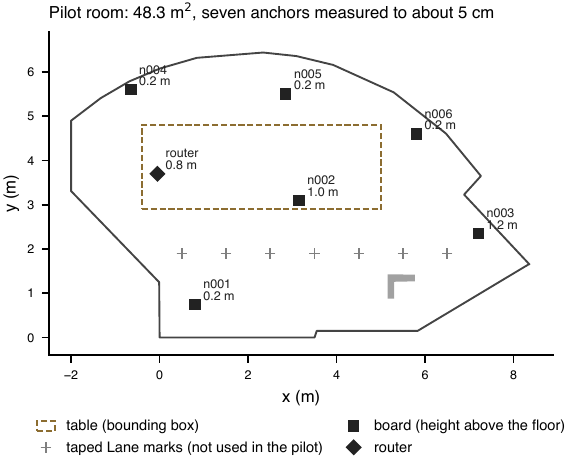}
\caption{The pilot room, drawn from the configuration snapshots of the session (room model, anchor positions and marks; no recording): the 19-vertex floor polygon from the scan, the wall stub, the table's bounding box, the seven anchors with their heights above the floor, and the taped Lane marks of the team's earlier ground-truth layout, which this session did not use (the boards were the marks).}
\label{fig:room}
\end{figure}

The session took place on 2026-10-11 in a room at the hackathon venue whose floor polygon
(19 vertices from the scan) covers 48.3~m$^2$ and spans 10.35 m by 6.44 m. A table stands along one
side; its bounding box is x from -0.4 to 5.0 m and y from 2.9 to 4.8 m. Table~\ref{tab:anchors} and Figure~\ref{fig:room} give
the anchor positions, measured to about 5 cm.

\begin{table}[H]
\centering
\caption{Anchor positions in the map frame (m), measured to about 5~cm; $z$ is the height above the floor.}
\label{tab:anchors}
\begin{tabular}{lrrr}
\toprule
Anchor & $x$ & $y$ & $z$ \\
\midrule
n001 & 0.80 & 0.75 & 0.2 \\
n002 & 3.15 & 3.10 & 1.0 \\
n003 & 7.20 & 2.35 & 1.2 \\
n004 & $-0.65$ & 5.60 & 0.2 \\
n005 & 2.85 & 5.50 & 0.2 \\
n006 & 5.80 & 4.60 & 0.2 \\
router & $-0.05$ & 3.70 & 0.8 \\
\bottomrule
\end{tabular}
\end{table}

\subsection{Protocol}

A script spoke the cues and logged their times to the second. The sequence was:

\begin{enumerate}
\def\labelenumi{\arabic{enumi}.}
\item
  Three minutes of empty room, door closed (15:53:30 to 15:56:30 local time).
\item
  Seven stops: on ``Go to n001'' the walker walks to the board, and on ``Stand still'' stands
  beside it for 15 s; then the next board, in the order n001, n002, n003, n006, n005, n004,
  router. From each ``Go to'' cue to the next ``Stand still'' cue took 10 to 13 s; the stops phase
  lasted about 3.5 min (15:56:37 to 16:00:03).
\item
  Two minutes of free walking anywhere in the room (16:00:07 to 16:02:07).
\item
  Two minutes of empty room (16:02:38 to 16:04:38).
\end{enumerate}

The Baseline was the empty-room Hold-still of 20 s taken at 15:27 the same afternoon and saved
by name;
its empty-room check had reported 10.3\% of Links reading motion and no Track. The tracker was
capped at one simultaneous Track for this session. The recording ran from 15:53:26 to 16:04:43
(678 s) and holds 211,254 datagrams, about 312 per second across the 36 Links.

The boards served as the position marks: the walker stood beside each board, so the ground truth
of a stop is the board position plus an offset of about 0.3 to 0.5 m, and the ground truth of a
walk is the straight line between two boards. None of the walked lines crosses the table; the
walks to and from n002 touch its edge, and the last metre into the router reaches its end.

\subsection{Video synchronisation}

An iPhone filmed the whole session (1920 x 1080, 29.97 fps, 828.7 s) from the router and n004
end of the room, looking along +x. The spoken cues are on the soundtrack. Using only the video's
own creation time, the 16 cues land within a median 0.18 s of the cue log (range -0.68 to
+1.53 s, with the log at 1 s resolution), so the video and the tracker data are in sync without a
clapper. The video confirms that the walker was where the cues said. The camera's own end of the
room (n004 and the router) is out of frame, which matters for the free walk (Section~\ref{sec:limitations}).

\subsection{Scoring}\label{sec:scoring}

All scores are computed from the tracker's recorded output (\texttt{tracks.jsonl}) and the cue log by
the scripts in \path{analysis/2026-10-11-solo/}:

\begin{itemize}
\item
  Walking between boards: for every tracker update between the ``Go to'' cue and the next ``Stand
  still'' cue, the perpendicular distance from the Track to the segment between the two boards,
  and the share of updates that carry a Track (\texttt{score\_lines.py}).
\item
  Arrival and stops: the Track's distance from the board from 1 s before to 2 s after the ``Stand
  still'' cue, and, with the first 3 s of each stop skipped, the median Track position over the
  rest of the stop against the board position (\texttt{score\_solo.py}).
\item
  Empty room: the share of updates with any Track, and how many distinct Track ids
  appeared (\texttt{score\_solo.py}).
\item
  Impossible positions: the share of Track positions outside the room polygon or inside the
  table's bounding box, using the scanned room model (\texttt{score\_impossible.py}).
\item
  Node-to-node versus router-only Links (Section~\ref{sec:e1}): the recording's raw datagrams are
  replayed offline through the same pipeline twice, once as recorded and once with every CSI
  frame whose transmitter is a board dropped before the pipeline, so that only the six router
  Links remain; the saved Baseline is restricted to the same six Links. Both runs re-apply the
  recorded settings and Track cap and are scored by the same scripts (\texttt{e1\_replay.py},
  \texttt{e1\_score.py}). The replay is deterministic: two runs of the same condition give identical
  Tracks. \texttt{e1\_score.py} adds a coverage-aware measure, the share of all walk updates (untracked
  ones counted as misses) with a Track within 1 m of the walked line, and 95\% confidence
  intervals from 2000 block-bootstrap resamples of 1 s blocks of the walk windows (seed 0);
  blocks rather than single updates because the 10 Hz updates are autocorrelated.
\item
  Error against the cue-based reference trajectory (\texttt{e3\_lag\_error.py}): the walker's reference
  position at time t is taken at constant speed along the straight line from the previous board
  to the cued board between the ``Go to'' cue and the ``Stand still'' cue, and the error is the
  distance from the Track to that position; untracked updates are misses. This reference is an
  assumption, not an independent measurement: the cue log has 1 s resolution, the walker stood
  0.3 to 0.5 m beside each board, and a walker who arrived before the cue or paused was not at
  the interpolated position, so the figure can be above or below the true error. It counts lag
  along the path, which the perpendicular distance does not. CIs as above, plus a bootstrap over
  the six walks for the median.
\item
  RTI baseline (\texttt{e2\_rti.py}): the standard radio tomographic reconstruction on the same anchors,
  grid cells and scoring, fed with two kinds of per-Span evidence taken from the all-Links
  replay output at 10 Hz: the pipeline's own per-Span CSI Disturbance, which isolates the
  imaging and tracking stage, and classic RSS shadowing, the drop of each Link's RSSI below its
  median over the first empty-room period. Each Span weights the cells inside the ellipse with
  its anchors as foci and an excess path lambda, scaled by one over the square root of the Span
  length; the image is the Tikhonov solution with a first-difference smoothness prior; the
  evidence is smoothed over 0.5 s; the position is the centroid around the brightest cell,
  reported when the peak exceeds a threshold. The threshold is set on the first empty-room
  period either strictly (99th percentile of the peak) or matched to the Ghostline run's own
  false-Track share there, so that both methods are compared at the same false rate. Lambda
  and the regularisation weight were swept (0.03, 0.06, 0.12 m; 1, 3, 10) and the best setting
  on these same walks is reported, which favours the baseline.
\end{itemize}

The perpendicular distance ignores lag along the path: a Track that waits at the previous board
while the walker moves still scores well. It is an optimistic bound; the error against the
cue-based reference is reported beside it. Neither is an independent position measurement;
annotating the walker in the video would supply one and is listed in Section~\ref{sec:planned}.

\section{Results}\label{sec:results}

\subsection{Link census}\label{sec:census}

During the session all six boards delivered CSI and all 36 directed Links were live, at about
312 datagrams per second in total. The pre-session check flagged three Links as slow (under
3 frames per second): n002 to n005, n002 to n006 and n006 to n002. A four-minute recording from
the previous evening in the same room, replayed by the team, also showed every Link live
throughout; that replay is an observation reported by the team and is not reproduced by the
scripts of this paper. Link uptime over hours is not measured.

\subsection{Walking between boards}

\begin{figure}[H]
\centering
\includegraphics[width=1.0\textwidth]{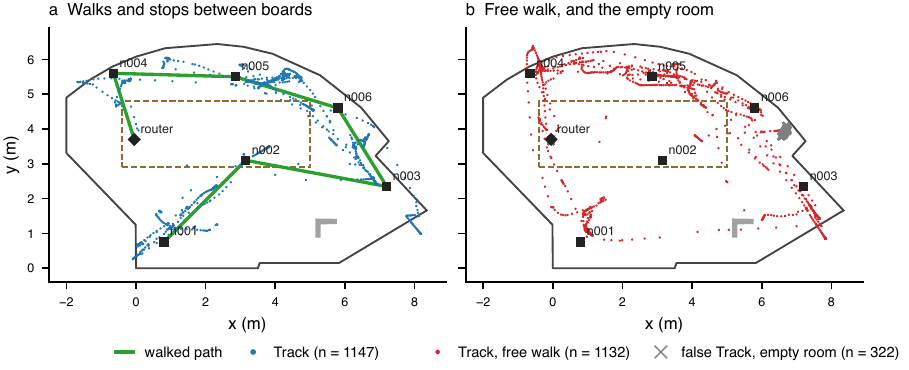}
\caption{Every Track position of the scored session \texttt{solo-empty-boards-20261011-1552} on the room outline, from the live \texttt{tracks.jsonl} and the cue log (\texttt{fig\_tracks\_phases.py}). (a) The walks and stops between boards, with the walked lines; $n$ is the number of Track positions. (b) The two-minute free walk, and the Track positions of the two empty-room periods (five minutes), which all sit at one spot near n006.}
\label{fig:phases}
\end{figure}

\begin{figure}[H]
\centering
\includegraphics[width=0.7\textwidth]{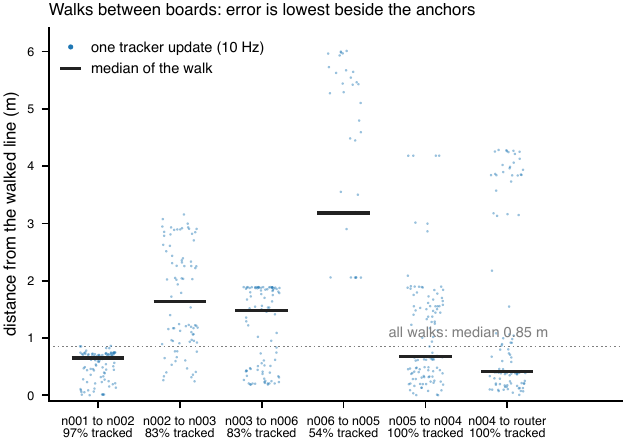}
\caption{Walks between boards in the scored session (\texttt{fig\_walk\_errors.py} on the live \texttt{tracks.jsonl}): for each walk, the perpendicular distance of every tracked update (10~Hz) from the straight line between the two boards, the walk's median, and below the axis the share of the walk's updates that carried a Track. The dotted line is the median over all six walks (553 tracked updates). The metric ignores lag along the line (Section~\ref{sec:scoring}).}
\label{fig:errors}
\end{figure}

Table~\ref{tab:walks} and Figures~\ref{fig:phases} and~\ref{fig:errors} give the result. Over the six walks, 644 tracker updates of which
553 carried a Track, the Track lay a median 0.85 m from the straight path, with a 90th percentile
of 3.18 m. Against the cue-based reference trajectory (Section~\ref{sec:scoring}) the median error is 1.56 m
(95\% CI 1.29 to 2.83 over 1 s blocks, 1.29 to 3.60 over walks), the 90th percentile 4.18 m, and
45.7\% of all walk updates (34.3 to 57.4) have a Track within 2 m of the reference, 21.3\% within
1 m. The perpendicular figure is about half of the reference-based one because it ignores lag;
neither is a measurement against observed positions.

\begin{table}[H]
\centering
\caption{Walks between boards: median perpendicular distance from the straight path, median error against the cue-based reference trajectory (Section~\ref{sec:scoring}), and share of updates with a Track.}
\label{tab:walks}
\small
\begin{tabular}{llll}
\toprule
Walk & Median off the path (m) & Median error vs. reference (m) & Tracked (\%) \\
\midrule
n001 to n002 & 0.65 & 1.20 & 97 \\
n002 to n003 & 1.64 & 2.23 & 83 \\
n003 to n006 & 1.48 & 2.49 & 83 \\
n006 to n005 & 3.18 & 4.03 & 54 \\
n005 to n004 & 0.68 & 3.22 & 100 \\
n004 to router & 0.42 & 1.31 & 100 \\
\midrule
All (553 tracked of 644) & 0.85 (90th pct 3.18) & 1.56 (90th pct 4.18) & 86 \\
\bottomrule
\end{tabular}
\end{table}

The walks n004 to router, n001 to n002 and n005 to n004 were the best (0.42 to 0.68 m); the
walk from n006 to n005 across the middle of the room was the worst and was tracked only 54\% of
the time. On n005 to
n004 the Track stayed near n005 while the walker moved and still scored 0.68 m, which is the
lag blind spot of the metric; against the cue-based reference its median error is 3.22 m.

\subsection{Arriving and standing still}

Arriving at a board (1 s before to 2 s after the ``Stand still'' cue), the Track was a median
1.74 m from the board, and 5 of 7 arrivals carried a Track. During the 15 s stops, with the
first 3 s skipped, a Track existed at only 4 of 7 boards, a median 3.18 m from the board. The
tracker follows movement, not presence: a person who stands still fades into the Baseline within
seconds, and whatever Track remains is noise. The standing-still row measures that behaviour,
not localisation accuracy.

\subsection{Free walk}

During the two-minute free walk a Track was present in 94\% of updates. Five Track ids were used,
so the walker was lost and re-acquired about four times. The free walk has no position ground
truth yet; the video can supply it (Section~\ref{sec:planned}).

\subsection{Empty room}

In the first three minutes of empty room a Track was present in 15.9\% of updates (two distinct
false Tracks); in the final two minutes, in 2.9\% (one). All false Tracks sat at about (6.7, 4.0),
near n006. Whatever moves there, a curtain, a door, a fan or a Link artefact, was not identified during
the session.

\subsection{Physically impossible positions}

During the cued period (the walks between boards and the stops at them), 4.1\% of Track
positions lay outside the room polygon and 7.7\% inside the table's bounding box; during the free
walk, 6.2\% and 6.6\%. None of the walked
lines crosses the table; the walks to and from n002 touch its edge, and the last metre into the
router reaches its end.

\subsection{Observations from the same afternoon, not scored}\label{sec:unscored}

Earlier the same afternoon, without ground truth, we watched the live display during informal
runs of 30 to 45 s each. These are observations, not scored results, and the data sit in the
black-box recordings without time-stamped ground truth:

\begin{itemize}
\item
  One walker walking freely, and one walker following voice cues, were each followed by a single
  Track along the route the walker reported, with one loss of a few seconds on the way back in
  the cued run.
\item
  Two walkers at once, in three runs: the tracker kept one Track that jumped between the
  walkers, and in the run with the Track cap verified at two a second Track appeared only
  briefly.
\item
  A second empty-room check, taken while the room was not in fact quiet, reported more than half
  of the Links reading motion and a Track in most updates, and was discarded. The contrast with
  the 10.3\% of the accepted check shows what the check is for.
\item
  The previous day's recordings, from a crowded room and an earlier layout with three receivers,
  show the busy-room limit: most Links were disturbed at once for long stretches, the grid had
  no quiet Links to vote against, and the tracker located nobody. These figures come from the
  team's logs and are not reproduced by the scripts of this paper.
\end{itemize}

\subsection{Node-to-node versus router-only Links}\label{sec:e1}

\begin{figure}[H]
\centering
\includegraphics[width=1.0\textwidth]{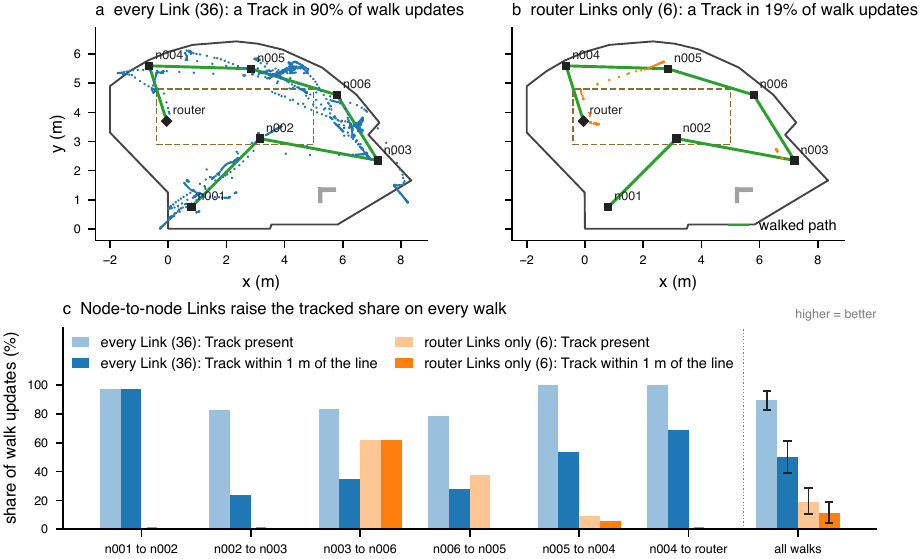}
\caption{The scored session replayed offline through the same pipeline (\texttt{fig\_e1.py} on the outputs of \texttt{e1\_replay.py} and \texttt{e1\_score.py}). (a, b) Track positions during the walks and stops with every Link (36) and with the router Links only (6), on the room outline with the walked path. (c) Per walk and over all six walks, the share of updates with a Track (light) and the share of all updates with a Track within 1~m of the walked line (dark), for both conditions; the error bars on the totals are 95\% CIs from 2000 block-bootstrap resamples of 1~s blocks. A short tick at zero marks a walk with no Track.}
\label{fig:e1}
\end{figure}

Table~\ref{tab:e1} and Figure~\ref{fig:e1} give the result of the offline replay (Section~\ref{sec:scoring}). Replayed as recorded,
with all 36 Links, the pipeline reproduces the live session closely: a Track in 90.0\% of the
660 walk updates against 85.9\% live, a median 0.83 m off the line against 0.85 m, the free walk
tracked in 96\% against 94\%, and false Tracks in 16.8\% and 3.1\% of the empty-room updates
against 15.9\% and 2.9\%. The live and replayed runs differ because the offline step is exactly
0.1 s of recording time while the live tick drifted (6,755 live updates against 6,778), and the
one walk where they differ by more than a few points is n006 to n005 (54\% live, 78\% replayed).

With the node-to-node Links removed and the six router Links left, on the same datagrams and
settings, a Track is present in 19.2\% of the walk updates (95\% CI 10.5 to 28.9, against 83.0 to
96.1 for all Links), none at all on three of the six walks, and the share of walk updates with a
Track within 1 m of the walked line falls from 50.0\% (39.1 to 60.9) to 11.2\% (4.5 to 18.8). The
median off the line over the updates that are tracked is lower, 0.34 m, because the 127 tracked
updates come mostly from one walk and from Tracks that sit near the table where the router
stands: 31.1\% of the router-only Track positions during the walks and stops lie inside the
table's bounding box, against 10.8\% with all Links. The free walk is tracked in 60\% of updates
with six Track ids, against 96\% and four. The router-only run produced no false Track in either
empty-room period, where the all-Links run produced them in 16.8\% and 3.1\% of updates: fewer
Links give fewer Tracks of both kinds.

\begin{table}[H]
\centering
\caption{The scored session replayed offline with every Link and with the router Links only. Walk windows as in Table~\ref{tab:walks}; 95\% CIs in brackets from the block bootstrap of Section~\ref{sec:scoring}; the live column is the recorded session for reference.}
\label{tab:e1}
\small
\begin{tabular}{>{\raggedright\arraybackslash}p{5.1cm}>{\raggedright\arraybackslash}p{2.9cm}>{\raggedright\arraybackslash}p{3.3cm}>{\raggedright\arraybackslash}p{3.3cm}}
\toprule
Measure & Live\newline (36 Links) & Replay,\newline every Link (36) & Replay,\newline router Links only (6) \\
\midrule
Walk updates & 644 & 660 & 660 \\
Updates with a Track (\%) & 85.9 [77.8, 92.9] & 90.0 [83.0, 96.1] & 19.2 [10.5, 28.9] \\
Median off the line, tracked updates (m) & 0.85 [0.66, 1.54] & 0.83 [0.67, 1.37] & 0.34 [0.34, 3.10] \\
90th percentile off the line (m) & 3.19 & 3.77 & 3.20 \\
Updates with a Track within 1~m of the line (\%) & 45.8 [34.8, 57.2] & 50.0 [39.1, 60.9] & 11.2 [4.5, 18.8] \\
Median error against the cue-based reference (m) & 1.56 [1.29, 2.83] & 1.65 [1.35, 2.74] & 0.93 [0.45, 4.64] \\
Updates with a Track within 2~m of the reference (\%) & 45.7 [34.3, 57.4] & 47.4 [36.4, 58.9] & 11.8 [4.5, 19.4] \\
Walks with no Track at all & 0 of 6 & 0 of 6 & 3 of 6 \\
Free walk: tracked (\%) / Track ids & 94 / 5 & 96 / 4 & 60 / 6 \\
Empty room, first 3 min: Track present (\%) / ids & 15.9 / 2 & 16.8 / 3 & 0.0 / 0 \\
Empty room, last 2 min: Track present (\%) / ids & 2.9 / 1 & 3.1 / 1 & 0.0 / 0 \\
Walks and stops: inside the table box (\%) & 7.7 & 10.8 & 31.1 \\
\bottomrule
\end{tabular}
\end{table}

This is one session, one room and one layout, replayed with settings that were tuned for the
full Link set. It is a dependence result, not a comparison of two tuned systems: a pipeline tuned
for six Links might do better than this router-only run, the full system also produced the
empty-room false Tracks that the router-only run did not, and the comparison says nothing about
other rooms. What it does show is that under the settings used the Tracks depended on the
node-to-node Links: without them the pipeline saw the walker in a fifth of the walk updates and
lost the middle and the far end of the room.

\subsection{A plain RTI image on the same evidence}\label{sec:rti}

\begin{figure}[H]
\centering
\includegraphics[width=0.72\textwidth]{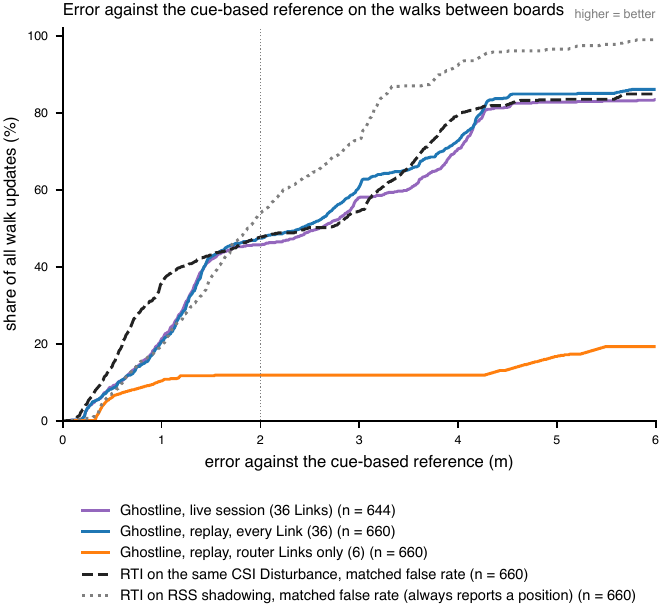}
\caption{Share of all walk updates with a position within a given error against the cue-based reference trajectory (\texttt{fig\_error\_cdf.py} on the per-update errors of \texttt{e3\_lag\_error.py} and \texttt{e2\_rti.py}), for the live session, the two replay conditions of Section~\ref{sec:e1} and the two matched RTI baselines of Section~\ref{sec:rti}. Untracked updates count as misses, so each curve's plateau is its tracked share; the RSS baseline reaches 100\% because it always reports a position. The dotted line marks 2~m.}
\label{fig:cdf}
\end{figure}

Table~\ref{tab:rti} and Figure~\ref{fig:cdf} give the RTI baseline (Section~\ref{sec:scoring}) beside the Ghostline grid and tracker
on the all-Links replay. Fed with the same per-Span CSI Disturbance and thresholded to the same
false-Track share in the first empty-room period (16.8\% of updates), the RTI image places a
position in 84.8\% of the walk updates against 90.0\%, with a median error against the cue-based
reference of 1.43 m against 1.65 m, 47.7\% of all walk updates within 2 m of the reference against
47.4\%, and 35.5\% within 1 m against 20.6\%; its false share in the last empty-room period is 4.2\%
against 3.1\%. Because both are scored on the same 660 updates, the differences carry paired
block-bootstrap intervals (\texttt{e2\_ci.py}; RTI minus Ghostline): tracked share -5.2 points (95\% CI
-15.3 to +4.4), within 2 m +0.3 points (-9.1 to +9.4), within 1 m +15.0 points (+7.0 to +23.3),
median error -0.23 m (-1.09 to +0.52). At the 2 m scale the two are indistinguishable within
about nine points either way, which is not evidence of equivalence; at the 1 m scale the RTI
image is ahead. The two curves in Figure~\ref{fig:cdf} lie on top of each other beyond 1.5 m. With the strict threshold the
RTI image reports a position in 41.5\% of the walk updates with a median error of 0.98 m and a
1.0\% and 0.6\% false share: a different operating point, not a better method. Classic RTI on RSS
shadowing, at the matched threshold, reaches 53.9\% within 2 m with a median of 1.88 m, but its
quiet reference drifted within the session: it reported a position in 100\% of the updates of the
last empty-room period, so the matched false rate holds only for the period it was set on.

\begin{table}[H]
\centering
\caption{RTI baseline against the Ghostline grid and tracker on the all-Links replay (660 walk updates); thresholds set on the first empty-room period as described in Section~\ref{sec:scoring}; $\lambda = 0.12$~m, regularisation weight 10, the best of the sweep for the CSI input. The last two rows give the difference to the Ghostline column with its paired block-bootstrap 95\% CI.}
\label{tab:rti}
\small
\begin{tabular}{>{\raggedright\arraybackslash}p{4.3cm}>{\raggedright\arraybackslash}p{2.3cm}>{\raggedright\arraybackslash}p{2.5cm}>{\raggedright\arraybackslash}p{2.5cm}>{\raggedright\arraybackslash}p{2.5cm}}
\toprule
Measure & Ghostline grid and tracker & RTI, CSI Disturbance, matched & RTI, CSI Disturbance, strict & RTI, RSS shadowing, matched \\
\midrule
Walk updates with a position (\%) & 90.0 & 84.8 & 41.5 & 100.0 \\
Median perpendicular distance to the line (m) & 0.83 & 0.36 & 0.21 & 1.04 \\
Median error against the cue-based reference (m) & 1.65 & 1.43 & 0.98 & 1.88 \\
Within 1~m of the reference, all walk updates (\%) & 20.6 & 35.5 & 21.1 & 19.2 \\
Within 2~m of the reference, all walk updates (\%) & 47.4 & 47.7 & 28.2 & 53.9 \\
Free walk: position present (\%) & 96 & 93 & 50 & 100 \\
Empty room, first 3 min: position present (\%) & 16.8 & 16.8 & 1.0 & 16.8 \\
Empty room, last 2 min: position present (\%) & 3.1 & 4.2 & 0.6 & 100.0 \\
\midrule
Difference within 1~m (points) [95\% CI] & & $+15.0$ [$+7.0$, $+23.3$] & $+0.6$ [$-6.4$, $+7.0$] & $-1.2$ [$-11.8$, $+9.5$] \\
Difference within 2~m (points) [95\% CI] & & $+0.3$ [$-9.1$, $+9.4$] & $-19.2$ [$-30.9$, $-7.6$] & $+6.5$ [$-11.1$, $+24.9$] \\
\bottomrule
\end{tabular}
\end{table}

The RTI image on the CSI Disturbance shares the first five stages of the pipeline (features,
motion statistic, Baseline, room level, Disturbance) and differs only in how the Spans are turned
into a position. On this recording that difference buys no measurable accuracy: at 2 m the two
cannot be told apart, and at 1 m the plain image is ahead. What the Ghostline grid, its
explain-away detector and the Kalman tracker add here is Track identities and coasting. We do
not claim equivalence of the two, and the attribution is limited to this recording: one walker,
six short walks, and a baseline whose two parameters were chosen on the same walks. Taken with
Section~\ref{sec:e1}, the observed pattern is that the Tracks depended on the Link set and not, as far as
this recording shows, on the imaging and tracking stage.

\section{What the numbers mean}\label{sec:discussion}

One person walking is tracked at the metre scale in this room, best along the side where three
anchors stand within a few metres of the path and worst across the middle. The map in Figure~\ref{fig:phases}
shows the Track positions hugging the anchors and the walls: evidence exists where Links cross,
and the middle of the room has few crossings in this layout. That is the geometric reason to
expect node-to-node Links to help, and the replay of Section~\ref{sec:e1} agrees with it for this layout:
the six router Links all start at the table edge, and without the crossing Links the pipeline
tracked the walker in a fifth of the walk updates and not at all on three of the six walks. The
replay is a controlled comparison on one recording, not a measurement of how much Links of
either kind are worth in general. The RTI comparison of Section~\ref{sec:rti} points the same way: a plain regularised image on the same
per-Span Disturbance tracks the walker as well as the grid and tracker at 2 m and better at 1 m,
so on this recording the last three stages buy Track identities and coasting rather than
accuracy, and the accuracy is set upstream, by the Links and the per-Link evidence.

The failures are structural, not tuning accidents. Standing still is invisible to a method that
compares each Link with its own recent past; presence detection needs a different statistic or
a longer memory. Two walkers were not separated with these anchors and this detector, and the
cap of one Track during the scored session means the false-Track counts are a lower bound. The
empty-room false Tracks sit at one spot, which points at a source in the room or at one Link
rather than at the pipeline as a whole, but the source is unknown.

Every threshold and parameter was tuned on synthetic data during the hackathon and checked by
eye in this room on the same afternoon. The session therefore does not separate tuning from
testing, and the numbers should be read as a demonstration that the pipeline works end to end
on hardware, not as its accuracy. The setting was not a laboratory: six modified consumer boards
and a home router installed in a day in a hackathon venue, with furniture, a table in the sensing
area and people passing outside the room; the numbers describe that setting, and the paper makes
no claim beyond it.

\section{Limitations}\label{sec:limitations}

\begin{itemize}
\item
  One walker, one room, one session: about a minute of walking with known positions, 11 minutes
  in all. Nothing here generalises to other rooms, layouts or people.
\item
  Ground truth is the board positions and the straight lines between them, with the walker's
  0.3 to 0.5 m offset from each board inside the error. The free walk has no position ground
  truth.
\item
  Neither error metric is a measurement against observed positions. The perpendicular distance
  ignores lag along the path and is an optimistic bound; the cue-based reference assumes constant
  speed between two cues at 1 s resolution and a walker 0.3 to 0.5 m beside each board, and can
  be above or below the true error. The video can supply observed positions for the part of the
  room it covers; that annotation has not been done.
\item
  The node-to-node versus router-only comparison is a replay of one session with settings tuned
  for the full Link set, not a comparison of two tuned systems, and the full system also produced
  the false Tracks. The RTI baseline on the CSI Disturbance shares the pipeline's first five
  stages, its two parameters were chosen on the same walks, and similar shares at one threshold
  do not establish equivalence; the RSS-based RTI is independent, but its quiet reference drifted
  within the session.
\item
  Settings were tuned on synthetic data and adjusted in the same room on the same day; there is
  no held-out session.
\item
  The tracker was capped at one Track in the scored session, so counting and false-Track
  figures are bounds, not rates per hour.
\item
  The pipeline is described in full with the parameter values used, but the code is not yet
  public, so reproducing the pipeline from the text means re-implementing it; the recordings and
  the scoring scripts reproduce every number in Section~\ref{sec:results}.
\item
  The system was built in three days by a hackathon team; the evaluation was designed and run
  by one of its members, not by an independent party.
\end{itemize}

\section{Further work}\label{sec:planned}

The recordings already made allow the following offline:

\begin{enumerate}
\def\labelenumi{\arabic{enumi}.}
\item
  Position ground truth from the video, for the free walk and for the walks between boards
  (the walker marked every 2 s where the camera covers the room), which would replace the
  cue-based reference with observed positions.
\item
  False Tracks per hour and Link uptime over hours from the two days of black-box recordings,
  with the Track cap lifted in the replay.
\item
  The node-to-node versus router-only replay of Section~\ref{sec:e1} and the RTI comparison of
  Section~\ref{sec:rti} on the previous evening's recording and on every further session; a router-only
  run with settings tuned for six Links; and an RTI baseline with its own Baseline and tracking
  stage rather than the pipeline's.
\end{enumerate}

A further session will add more laps, a second walker, two walkers at once, a crowd, one node
unplugged, and if possible a second room, with settings frozen before the session.

\section{Conclusion}

Ghostline shows that six ESP32-S3 boards and a consumer router can sustain 36 directed CSI
Links in a room and that a signal-processing pipeline on a laptop can turn them into a Track
that follows one walker at the metre scale, with a protocol that ties every tracker output to a
synchronised video. It also shows, in the same session, that the system loses a person who
stands still and produces a false Track at one spot in an empty room, and, in the informal runs
of the same afternoon, that it does not separate two walkers in this layout. Replayed with the node-to-node Links removed, the same pipeline on the
same recording tracked the walker in a fifth of the walk updates instead of nine tenths, which
is what the link set was built for, and a plain RTI image on the same per-Span evidence did as
well as the grid and tracker at 2 m and better at 1 m, which locates the measured dependence in
the Links rather than in the imaging and tracking stage. How far that holds beyond one room, one
layout and one set of settings is the question the next sessions answer.

\section*{Data and code availability}

The recording of the scored session (raw datagrams, tracker output, events and metadata), the
afternoon's black-box recordings, the cue log, the configuration snapshots, the two replay
outputs of Sections~\ref{sec:e1} and~\ref{sec:rti}, the scored results and the analysis scripts are deposited on
Zenodo (DOI 10.5281/zenodo.23337813). From the deposited files the scripts recompute every number
and figure of this paper. The replay itself, which turns the raw datagrams into the two replay
outputs, needs the tracker code, which is in the team's private repository and whose release has
not been decided; the method is described in full in Section~\ref{sec:system}. The video, which shows the
walker, is available from the corresponding author on request.

\section*{Ethics}

The system tracks anonymous motion: a Track is a position with a number, and no image, identity
or device is captured. The only person tracked in the scored session was one of the authors (S.C.),
who designed the session and walked it. The same afternoon's black-box recordings (Section~\ref{sec:unscored})
were made in the team's room with only the authors and the acknowledged team members in range of
the sensor network, all of whom knew the system was recording and agreed to it. The previous
day's figures on a crowded room come from the team's own logs of its demonstration, during which
the people present were told that an anonymous Wi-Fi motion demonstration was running and
recording, and no one objected; those logs are not deposited and are reported only as aggregate
Link statistics and as the tracker locating nobody, with no position of any person.

\section*{Acknowledgements}

Andrei Pax Popa validated and diagnosed the boards and wrote the React diagnostics app, and
Andra Tritza built the pitch deck variants, both during the hackathon; neither part is described
in this paper. We thank the organisers of the European Defense Tech Hackathon in Bucharest for the venue
and the occasion.

Contributions. Ionuț Moldovanu built the hardware: the boards, their mounting and the sensor
network. Alexandru Constantin wrote the software: the firmware modifications, the relay, the
tracker (pipeline, room check, recording and replay), the browser page and the venue model in the
map frame. Sebastian Cochinescu scanned the rooms and merged the scans, designed and ran the
evaluation (the cued-walk protocol, the cue script, the video and the scoring), was the walker in
the scored session, managed the project and wrote the paper.

\section*{Tool-use disclosure}

Large language models (Anthropic, OpenAI) were used as drafting and revision aids and to write the
deposited analysis scripts. They did not collect the data or run the session. The authors directed
and reviewed their use and take responsibility for the manuscript.

\bibliographystyle{plain}
\bibliography{references}

\end{document}